\documentstyle[11pt]{article}
\newcommand{\be}{\begin{equation}}
\newcommand{\ee}{\end{equation}}
\newcommand{\bea}{\begin{eqnarray}}
\newcommand{\eea}{\end{eqnarray}}
\newcommand{\intl}{\int_{-L}^{L}}

\newcommand{\AmS}{{\protect\the\textfont2
  A\kern-.1667em\lower.5ex\hbox{M}\kern-.125emS}}

\begin{document}

\rightline{\large TPR 94-24}
\rightline{hep-th/9409084}

\vspace*{.5cm}

\centerline{\huge \bf Light Front Quantisation of Gauge Theories}
\centerline{\huge   \bf   in  a  Finite   Volume{\parindent   0pt
\footnote{Talk  presented  at QCD 94, Montpellier,  France,  July
7-13,  1994, and at the Fourth  International  Workshop  on Light
Cone   Quantization   and   Non-Perturbative   Dynamics,   Polana
Zgorzelisko, Poland, August 15-25, 1994}}}

\vspace*{1cm}

\noindent
T.~Heinzl,  Institut  f\"ur  Theoretische  Physik,  Universit\"at
Regensburg,    93040   Regensburg,   Germany   {\parindent    0pt
\footnote{E-mail: Thomas.Heinzl@physik.uni-regensburg.de}}

\vspace*{.5cm}

\leftline{\bf Abstract:}
\noindent
We discuss  the  light front  formulation  of $SU(2)$  Yang-Mills
theory on a torus.

\section{Introduction}

In  1949, Dirac,  when  formulating  his  rules  of  relativistic
dynamics\cite{Dirac},  suggested  that one might use the variable
$x^+ = t + x^{3}/c$ as a new time parameter  instead of the usual
Galileian time~$t$.  This new formulation, replacing the standard
``instant form" of dynamics, he called ``front form" referring to
the fact that light fronts  (LFs) $x^+ = const$  are now surfaces
of equal time.  Time evolution  is accordingly  governed  by a LF
Hamiltonian  $P^- \equiv P^0 - P^3$. In the late sixties, Dirac's
idea  was  extended  to field  theory  in the context  of current
algebra and the quark parton model.  This development  culminated
in the formulation of LF QCD a decade later\cite{LFQCD}.   \par
The basic ingredient  of canonical LF QCD is the use of the light
cone (LC) gauge $A^+ \equiv  A_{-} = 0$, which is physical  (i.e.
free of ghosts and negative  metric states) and leads to a simple
elimination  of redundant degrees of freedom.  The LF Hamiltonian
becomes thus a function of the transverse  gauge potentials $A_i$
and the ``good"  fermion  components  $\psi_+$  only.   The other
important  feature of canonical  LF QCD is the triviality  of the
vacuum, which allows for a Fock expansion  of hadrons in terms of
only  a  few  constituents.   This  has  led  to  the  method  of
discretised light cone quantisation (DLCQ)\cite{DLCQ},  which has
been used to solve (low-dimensional)  quantum field theories on a
longitudinal momentum lattice.\par
In the meantime,  however, it has become clear that the canonical
formulation of LF QCD suffers from several problems: the LC gauge
shares  all the troubles  known from axial gauges,  in particular
infrared  singularities,  $1/(k^+)^n$,  related to residual gauge
invariance.  These induce symmmetry violating counterterms  which
obscure the renormalisability of the theory.  Furthermore, it has
become apparent  that the vacuum is trivial  only if there are no
dynamical  modes present with LF momenta  $k^+ = {\bf k_\perp}  =
0$, called zero modes (ZMs)\cite{vacuum}.\par
To regulate  and resolve the infrared  problems we will work in a
finite spatial  volume,  which allows a clear separation  between
ZMs and normal modes.  Our final goal is to find a LF Hamiltonian
with  the  ZMs included  that  might  serve  as a basis  for DLCQ
calculations.  This contribution describes the first few steps of
this program.

\section{The LC Gauge on a Torus}

We compactify by the restriction $-L \le x^-$, $x_1$, $x_2 \le L$
and  by imposing  periodic  boundary  conditions  for  the  gauge
fields.  Space thus topologically  becomes a torus.  Consider the
Wilson loop winding around the torus in $x^-$-direction,

\be
W[A]= \frac{1}{N}  \hbox{tr  P} \exp \left[-i  g \intl dx^- A_{-}
\right] .
\ee

\noindent
Choosing the LC gauge $A_- = 0$ amounts to prescribing  the value
$W[A] =1$ for this gauge invariant  dynamical  quantity  which is
clearly forbidden.  The LC gauge on a torus, therefore, cannot be
attained.   A possible gauge choice, however, is $A_{-} = Y ({\bf
x})$,  where  a  ZM  $Y({\bf   x})$  with  respect  to  $x^-$  is
retained\cite{Yabuki}.

\section{Complete Abelian Gauge Fixing}

As a warm-up excercise we treat pure electromagnetism  coupled to
external sources $J_{\mu}$ in $d = 3+1$ including all ZMs.  First
we need some notation.   We write  any phase  space variable  $f$
(gauge  potentials  and  conjugate  momenta)  as  a sum  of  four
components  $f_r$, $f = f_0 + f_1 + f_2 + f_3$, which are defined
as

\bea
f_3 & \equiv  & f (x) - \frac{1}{2L}  \int \! dx_3 \,
f(x) \nonumber \\
f_2  & \equiv  & \frac{1}{2L}  \int \!  dx_3 \, f(x)  -
\frac{1}{(2L)^2} \int \! dx_3 dx_2 \, f(x) \nonumber \\
f_1  & \equiv  & \frac{1}{(2L)^2}  \int \! dx_3  dx_2 \, f(x)  -
\frac{1}{(2L)^3} \int \! d^3x \,f(x) \nonumber \\
f_0 & \equiv & \frac{1}{(2L)^3} \int \! d^3x \,f(x) \nonumber \; .
\eea

\noindent
We have denoted  $x_3 \equiv x^-$; all integrations  extend  from
$-L$ to $L$.  The index $r$ counts on how many spatial  variables
$f_r$ depends; $f_0$, for example, is thus constant  in all three
spatial directions (a global ZM). The components $f_r$ obey

\be
\intl dx_r f_r = 0 \; ,
\ee

\noindent
which states the absence of a ZM with respect  to $x_r$ in $f_r$.
\par
Before any elimination  of variables the canonical  pairs (in the
Poisson  bracket  sense)  are $(A_-, \Pi_+) \equiv (A_3, \Pi_3)$,
$(A_i, \Pi_i)$; $i= 1,2$. After decomposition, Gauss's law reads

\bea
\partial_3  \Pi_{33} + \partial_2  \Pi_{23} + \partial_1 \Pi_{13}
& = & J_{-3}  \nonumber\\
\partial_2 \Pi_{22} + \partial_1 \Pi_{12} & = & J_{-2} \nonumber\\
\partial_1 \Pi_{11} & = & J_{-1} \nonumber\\
0 & = & J_{-0} \; .
\eea

\noindent
The components $\Pi_{rr}$, $r = 1,2,3$ are to be eliminated. This
can be done without ambiguities due to condition (2). The inverse
operators   $\partial_r^{-1}$   are   given   by  periodic   sign
functions\cite{HW}.  \par
A natural  gauge  choice  for the case at hand is suggested  by a
method  due to Faddeev  and Jackiw\cite{FJ}:   set to zero  those
components of the gauge fields the conjugate momenta of which are
eliminated  via Gauss's  law.  Here, this leads to a gauge fixing
first introduced by Palumbo in order to modify the axial gauge on
a torus\cite{Palumbo},

\be
A_{33} = A_{22} = A_{11} = 0 \; .
\ee

\noindent
This gauge will henceforth  be called Palumbo  gauge.  Its actual
virtues  will become  more clear in the non-abelian  case.  There
still is a residual gauge freedom consisting  of discrete  shifts
of  the  global  ZMs  $A_{r0}$.    This  can  be  eliminated   by
appropriately  restricting  the range of $A_{r0}$ and will not be
discussed  here.   \par
After solving additional,  LF specific,  second class constraints
the Hamiltonian on the reduced phase space becomes

\bea {\cal H}_{red} & = & - \frac{1}{2} A_{i3}\Delta_\perp A_{i3}
- (\partial_i A_{i3}) (\partial_3^{-1}  J_{-3}) -  J_{i3}A_{i3}
- \frac{1}{2}  J_{-3}  \partial_3^{-2}  J_{-3}  + \nonumber\\
& + & \frac{1}{2}  \Pi_{30}^2  -  J_{+0}  A_{30} -
J_{+2}  \Delta_\perp^{-1}   J_{-2}  +
\frac{1}{2} J_{12} \Delta_{\perp}^{-1}  J_{12} - J_{+1}
\partial_1^{-2}     J_{-1}     +     \frac{1}{2}
J_{21}\partial_1^{-2} J_{21} \; . \eea

\noindent
Note the appearance of additional Coulomb terms induced by ZMs in
the last line.  Eq.~(5) should be compared with the naive version
(without ZMs) given by

\be
{\cal H}'_{red}   =  - \frac{1}{2}  A_{i}\Delta_\perp
A_{i} - (\partial_i A_{i}) (\partial_3^{-1} J_{-}) -
J_{i}A_{i} - \frac{1}{2} J_{-} \partial_3^{-2}  J_{-} \; ,
\ee

\noindent
where  some  ad  hoc  prescription   of  the  inverse  derivative
$\partial_3^{-1,2}$ has to be used.

\section{Complete Non-Abelian Gauge Fixing}

We consider  the  simplest  example,  namely  $SU(2)$  Yang-Mills
theory.  We mainly use matrix notation  $f \equiv f^a \tau^a /2$;
$a = 1,2,3$, the $\tau^a$ being the standard Pauli matrices.  The
canonical  pairs  before  any elimination  are the same as in the
abelian case.  With the help of the covariant  derivative  $D_r =
\partial_r  + ig [A_r, \quad] $; $r = 1,2,3$, Gauss's  law can be
written as

\be
G \equiv \sum_{r=1}^3 D_r \Pi_r = 0 \; .
\ee

Again  we want  to solve  for the projections  $\Pi_{rr}$;  $ r =
1,2,3$,  and the corresponding  gauge choice is the Palumbo gauge
$A_{rr}  = 0$.   If we decompose  Gauss's  law  according  to the
different  phase space sectors we first find the global ZM $G_0 =
0$.  This constraint  will not be discussed  here.   In the other
phase space sectors ($r > 0$) we obtain

\be
{\cal D}_r \Pi_{rr} \equiv \partial_r \Pi_{rr} + ig [{\cal A}_r ,
\Pi_{rr} ] = R_r \; ,
\ee

\noindent
where the field ${\cal A}_r = \sum_{s=0}^{r-1}A_{rs}$  is the sum
of all ZMs of $A_r$ with respect to $x_r$ and therefore  does not
depend  on $x_r$ which will become  important  in a moment.   The
inhomogeneity $R_r$ is independent of $\Pi_{rr}$. Thus, if we can
invert the covariant  derivative  ${\cal D}_r$ (which essentially
is the Faddeev-Popov  (FP) matrix),  we can solve Gauss's law for
$\Pi_{rr}$.   Note that in contrast  to the naive LC gauge the FP
matrix  depends  on the gauge  fields,  but only  on the  special
configurations  ${\cal  A}_r$.  As these are ZMs with respect  to
$x_r$, i.e.  $\partial_r  {\cal A}_r = 0$, and as ${\cal D}_r$ is
an  ordinary  (and  not  a partial)  differential  operator,  the
eigenvalue  problem of the covariant derivative  factorises  into
space and color and becomes exactly solvable!  This means that we
can determine the spectral decomposition  of the associated Green
function $G_{(r)}$ and solve for the momenta $\Pi_{rr}  = G_{(r)}
* R_r$.   For the axial  gauge  this has been  already  noted  by
Palumbo\cite{Palumbo}.

The eigenfunctions  $u_{n_r , \alpha_r}$  are a product  of plane
waves  $\exp(i  \pi  n_r  x_r  /L)$  and  color  matrices.    The
eigenvalues $\lambda_{n_r, \alpha_r}$ are given by

\be
\lambda_{n_r , \alpha_r} = \frac{\pi n_r}{L} + g \alpha_r \; .
\ee

\noindent
Here, the numbers $n_r$ are integer and $\alpha_r  = 0, \pm \vert
{\cal  A}_r \vert$,  where $\vert {\cal A}_r \vert \equiv  ({\cal
A}_r^a{\cal  A}_r^a)^{(1/2)}$  is the magnitude  of the isovector
with components  ${\cal A}_r^a$.  Thus, there would be ZMs of the
FP matrix if $\vert {\cal A}_r \vert = \pi n_r/gL$.  However,  if
we restrict the range of ${\cal A}_r$ to the fundamental  modular
domain\cite{vanBaal}, $ \vert {\cal A}_r \vert < \pi n_r/gL$, the
ZMs are absent  and we can uniquely  invert  ${\cal  D}_r$  which
yields the Green function

\be
G_{(r)}     (x_r     ,     y_r)     =    \sum_{n_r     \ne     0;
\alpha_r}\frac{u^{\dagger}_{n_r,    \alpha_r}(x_r)    u_{n_r    ,
\alpha_r} (y_r)}{ \pi n_r/L + g \alpha_r} \; .
\ee

\section{Conclusions}

As the LC gauge  $A_-  = 0$ does  not  exist  on a torus  we have
chosen  a modification  due to Palumbo  which  preserves  as many
virtues and avoids as many problems  of the LC gauge as possible.
The new gauge leads to a field dependent  FP matrix; nevertheless
it allows  for  an exact,  non-perturbative  solution  of Gauss's
law,  a property  not shared  by e.g.~the  Coulomb  gauge.   What
remains  to  be done  is to solve  the  residual  (second  class)
constraints   and  eventually   obtain   the  finite   volume  LF
Hamiltonian as an input for actual DLCQ calculations.

\vglue .5truecm

\leftline{\bf Acknowledgements}

\vglue .3truecm
\noindent
The  author  thanks  the  organisers  of  both  conferences,   in
particular S.~Narison and S.~G\l azek, for their efforts.

\vglue 0.5truecm

\end{document}